\documentclass[10pt]{article}
\newcommand{\be}{\begin{equation}}
\newcommand{\ee}{\end{equation}}
\newcommand{\nbar}[1]{\overline{#1}}                       % big bar
\def\bea{\begin{eqnarray}}
\def\eea{\end{eqnarray}}
\def\beas{\begin{eqnarray*}}
\def\eeas{\end{eqnarray*}}

\def\parp{{\partial}^{+}}
\usepackage{latexsym}

\begin{document}
\begin{titlepage}
\begin{flushright}    UFIFT-HEP-04-08 \\ 
%hep-ph/0405150 
\end{flushright}
\vskip 1cm
\centerline{\LARGE{\bf {Oxidizing  SuperYang-Mills from  }}}
\vskip .5cm
\centerline{\LARGE{\bf { $(N=4\ ,d=4)$ to $(N=1\ ,d=10)$ }}}

\vskip 1.5cm
\centerline{\bf Sudarshan Ananth\footnote{Supported by an Alumni Fellowship from the University of Florida}, } 
\vskip .5cm
\centerline{\em  Institute for Fundamental Theory,}
\centerline{\em Department of Physics, University of Florida}
\centerline{\em Gainesville FL 32611, USA}
\vskip .5cm
\centerline{\bf Lars Brink,  }
\vskip .5cm
\centerline{\em Department of Theoretical Physics}
\centerline{\em Chalmers University
of Technology, }
\centerline{\em S-412 96 G\"oteborg, Sweden}

\vskip .5cm
\centerline{\bf Pierre Ramond${}^{\,}$\footnote{Supported in part
by the US Department of Energy under grant DE-FG02-97ER41029} }
\vskip .5cm
\centerline{\em  Institute for Fundamental Theory,}
\centerline{\em Department of Physics, University of Florida}
\centerline{\em Gainesville FL 32611, USA}

\vskip 1.5cm

\centerline{\bf {Abstract}}
\vskip .5cm
\noindent  We introduce  superspace generalizations of the transverse derivatives to rewrite  the four-dimensional $N=4$ Yang-Mills theory into the  fully ten-dimensional $N=1$ Yang-Mills in light-cone form. The explicit SuperPoincar\'e algebra is constructed and invariance of the ten-dimensional action is proved. 

\vfill
\begin{flushleft}
May 2004 \\
\end{flushleft}
\end{titlepage}

\section{Introduction}
Superstring theories  are tantalizingly close to  the real world, in the sense that they include both gravitational and Yang-Mills interactions, but  they suffer from too many symmetries and supersymmetries, and appear content to stay in ten dimensions. Comparison with data requires an understanding, perhaps dynamical, of the breaking of these  symmetries, for which we have no hint.

One ray of hope is  afforded by the eleven-dimensional M-theory\cite{WITTEN}. Although we know very little about it, save for its existence, it  subsumes  all known superstring theories, and may contain within it guideposts for supersymmetry-breaking and dimensional  reduction. Its infrared limit is the much studied $N=1$ supergravity in eleven dimensions\cite{N=1},  believed to be ultraviolet divergent. 

Tracking  and fixing divergences has proven to be a fruitful approach to physics, and   fixing the divergences of $N=1$ supergravity may well be a key to a deeper understanding of $M$-theory.

A technically difficult but conceptually simple framework for discussing divergences is the light-cone\cite{DIRAC} frame formulation. In ten dimensions, the lack of divergences in string theories can be attributed to the triality of $SO(8)$, the light-cone little group.  The vector supermultiplet in ten dimensions with eight bosons and eight fermions produces, when dimensionally reduced, the $N=4$ SuperYang-Mills theory in four dimensions, which is free of ultraviolet divergences\cite{SWEDES}\cite{MANDELSTAM}. In 1982 Curtright\cite{CURTRIGHT} had in fact conjectured such a group theoretical relation between ultraviolet divergences and transverse little group representations for dimensionally-reduced theories.  He applied a similar reasoning to eleven dimensions,  and conjectured that the incomplete cancellation of the Dynkin indices of the $SO(9)$ representations that describe $N=1$ supergravity  was responsible for the divergences of that theory.

Interestingly, the relations among the Dynkin indices of the three representations of $N=1$ Supergravity in eleven dimensions could be generalized to an infinite set of triplets of $SO(9)$ representations,  with the same incomplete cancellation among the eighth-order Dynkin indices\cite{PENGPAN}. Their mathematical origin was elucidated\cite{GKRS}, and they were expressed in terms of light-cone superfields\cite{BRX}. They describe  massless fields of spin greater than two. Only with an infinite number of these Euler triplets can  the well-known no-go theorems be evaded, pointing to a non-local generalization of $N=1$ supergravity.  To make this problem tractable,  we first need to develop a  light-cone description of $N=1$ supergravity in eleven dimensions, and study its divergences.  We intend to extend these calculations to include the Euler triplets. 

This paper is a first step in this program. We study the simpler case of the SuperYang-Mills supermultiplet in ten dimensions, which is the massless sector of an open string theory. Starting from its dimensionally-reduced light-cone formulation, we show how to oxidize it to a fully ten-dimensional theory. Amazingly, the theory in four dimensions has so much memory of its original dimension  that this can be done by simply introducing  superspace generalizations of the transverse derivatives.  The  simplicity and compactness of the result makes us hopeful that it can be used to study all the symmetries of the ten-dimensional SuperYang-Mills theory. 

Of course, the ten-dimensional theory is no longer ultraviolet finite, unless one adds the massive string states (a world-sheet approach to field theory such as in ref. \cite{THORN} may be useful to develop this idea further). In future works, we aim to study how this taming of the field theory divergences comes about in the light-cone formalism.  This ten-dimensional formulation may also be a stepping stone for developing the corresponding formalism for  eleven-dimensional supergravity.

\section{The $LC_2$ Formulation of $N=4$}
Our starting point for this paper is the 1982 light-cone formulation of the $N=4$ SuperYang-Mills theory\cite{TEAM}, as originally described in ref  \cite{CHALMERS1}. This theory is the compactified version of  SuperYang-Mills in ten dimensions, with one vector and one spinor  field, both massless, linked by $N=1$ supersymmetry, and members of the adjoint representation of a  Lie algebra. In the unitary light-cone gauge with only physical degrees of freedom, they   transform as the eight-dimensional representations (one bosonic, one fermionic) of $SO(8)$, the transverse massless little group.

When this theory is dimensionally reduced to four dimensions, it consists of one complex bosonic field (the gauge field), four complex Grassmann fields and six scalars.   
\subsection{Field Content}
With the space-time metric $(-,+,+,\dots,+)$, the light-cone coordinates  and their derivatives are 

 \bea
{x^{\pm}}&=&\frac{1}{\sqrt 2}\,(\,{x^0}\,{\pm}\,{x^3}\,)\ ;\qquad ~ {\partial^{\pm}}=\frac{1}{\sqrt 2}\,(\,-\,{\partial_0}\,{\pm}\,{\partial_3}\,)\ ; \\
x &=&\frac{1}{\sqrt 2}\,(\,{x_1}\,+\,i\,{x_2}\,)\ ;\qquad  {\bar\partial} =\frac{1}{\sqrt 2}\,(\,{\partial_1}\,-\,i\,{\partial_2}\,)\ ; \\
{\bar x}& =&\frac{1}{\sqrt 2}\,(\,{x_1}\,-\,i\,{x_2}\,)\ ;\qquad  {\partial} =\frac{1}{\sqrt 2}\,(\,{\partial_1}\,+\,i\,{\partial_2}\,)\ ,
\eea
so that 
\be
{\parp}\,{x^-}={\partial^-}\,{x^+}\,=\,-\,1\ ;\qquad {\bar \partial}\,x\,=\,{\partial}\,{\bar x}\,=+1 \ .
\ee
In four dimensions,  any massless particle can be described by a  complex  field, and its complex conjugate of opposite helicity, 
the $SO(2)$ coming from the little group decomposition

\be
SO(8)\supset~SO(2)~\times~SO(6)\ .
\ee
Particles with no helicity are described by   real  fields.  The eight vectors fields in ten dimension reduce to

\be
{\bf 8}^{}_v~=~{\bf 6}^{}_0+{\bf 1}^{}_1+{\bf 1}^{}_{-1}\ ,
\ee
 and the eight spinors  to

\be
{\bf 8}^{}_s~=~{\bf 4}^{}_{1/2}+{\bf \bar4}^{}_{-1/2}\ .
\ee
The representations  on the right-hand side belong to $SO(6)\sim SU(4)$, with subscripts denoting the helicity: there are six scalar fields,  
 two vector fields, four spinor fields  and their conjugates. To describe them in a compact notation, we  introduce anticommuting Grassmann variables $\theta^m$ and $\bar\theta_m$, 

\be
\{\,{\theta^m}\,,\,{{ \theta}^n}\,\}\,~=~\,\{{{\bar \theta}_m}\,,\,{\bar\theta_n}\,\}\,~=~\,\{{{\bar \theta}_m}\,,\,{\theta^n}\,\}\,~=~0\ ,
\ee
which transform as the spinor representations of $SO(6)\sim SU(4)$,

\be
\theta^m_{}~\sim~ {\bf 4}^{}_{1/2}\ ;\qquad \overline\theta_{}^m~\sim~ {\bf \bar 4}^{}_{-1/2}\ ,
\ee
where $m,n,p,q,\dots =1,2,3,4$, denote $SU(4)$ spinor indices. Their  derivatives are written as 

\be
{{\bar \partial}_m}\,~\equiv~\,\frac{\partial}{\partial\,{\theta^m}}\ ;\qquad{\partial^m}\,~\equiv~\,\frac{\partial}{\partial\,{\bar \theta}_m}\ ,
\ee
with canonical anticommutation relations

\be
\{\,{\partial^m}\,,\,{{\bar \theta}_n}\,\}\,~=~\,{{\delta^m}_n}\ ;\qquad \{\,{{\bar \partial}_m}\,,\,{\theta^n}\,\}\,~=~\,{{\delta_m}^n}\ .
\ee
Under conjugation, upper and lower spinor indices are interchanged, so that $\nbar{{\theta^m}}\,=\,{{\bar \theta}_m}$, while  

\be 
\nbar{({\bar \partial}_m)}~=~-\partial^m\ ;\qquad\overline{(\partial^m)}~=~-\,{\bar \partial_m}\ .
\ee
Also,  the order of the operators is interchanged; that is $\nbar{{\theta^m}\theta^n}\,=\,{{\bar \theta}_n\,\bar\theta_m}$, and 
$ \nbar{\partial^m\,\partial^n}~=~{\bar \partial}_n\,{\bar \partial}_m$. 

 {\it All} the physical degrees of freedom  can be captured  in one complex superfield 

\bea
\phi\,(y)&=&\frac{1}{ \partial^+}\,A\,(y)\,+\,\frac{i}{\sqrt 2}\,{\theta_{}^m}\,{\theta_{}^n}\,{\nbar C^{}_{mn}}\,(y)\,+\,\frac{1}{12}\,{\theta_{}^m}\,{\theta_{}^n}\,{\theta_{}^p}\,{\theta_{}^q}\,{\epsilon_{mnpq}}\,{\partial^+}\,{\bar A}\,(y)\cr
& &~~~ +~\frac{i}{\partial^+}\,\theta^m_{}\,\bar\chi^{}_m(y)+\frac{\sqrt 2}{6}\theta^m_{}\,\theta^n_{}\,\theta^p_{}\,\epsilon^{}_{mnpq}\,\chi^q_{}(y) \ .
\eea
In this notation, the eight original gauge fields $A_i\ ,i=1,\dots,8$ appear as

\be
A~=~\frac{1}{\sqrt 2}\,(A^{}_1+i\,A^{}_2)\ ,\qquad \bar A~=~\frac{1}{\sqrt 2}\,(A^{}_1-i\,A^{}_2) \ ,
\ee
while the six scalar fields are written as antisymmetric $SU(4)$ bi-spinors 

\bea
C_{}^{m\,4}~=~\frac{1}{\sqrt 2}\,({A^{}_{m+3}}\,+\,i\,{A^{}_{m+6}})\ ,\qquad \nbar C_{}^{m\,4}~=~\frac{1}{\sqrt 2}\,({A^{}_{m+3}}\,-\,i\,{A^{}_{m+6}})\ ,
\eea
for $m\;\neq\,4$; complex conjugation is akin to duality, 
\bea
\label{dual}
{{\nbar C}^{}_{mn}}~=~\,\frac{1}{2}\,{\epsilon^{}_{mnpq}}\,{C_{}^{pq}} \ .
\eea
The fermion fields are denoted by $\chi^m$ and $\bar\chi_m$. All have adjoint indices (not shown here), and are local fields in the  modified light-cone coordinates  

 \bea
y~=~\,(\,x,\,{\bar x},\,{x^+},\,y^-_{}\equiv {x^-}-\,\frac{i}{\sqrt 2}\,{\theta_{}^m}\,{{\bar \theta}^{}_m}\,)\ .
\eea
In this particular light-cone formulation, called $LC_2$ by some,  all the unphysical degrees of freedom have been integrated out, leaving only  the physical ones.

Introduce the chiral derivatives,  

\bea
{d^{\,m}}=-{\partial^m}\,-\,\frac{i}{\sqrt 2}\,{\theta^m}\,{\partial^+}\ ;\qquad{{\bar d}_{\,n}}=\;\;\;{{\bar \partial}_n}\,+\,\frac{i}{\sqrt 2}\,{{\bar \theta}_n}\,{\partial^+}\ ,
\eea
which satisfy  the anticommutation relations

\be
\{\,{d^m}\,,\,{{\bar d}_n}\,\}\,=\,-i\,{\sqrt 2}\,{{\delta^m}_n}\,{\parp}\ .
\ee
One verifies  that $\phi$ and its complex conjugate $\bar\phi$ satisfy the chiral constraints

\be
{d^{\,m}}\,\phi\,=\,0\ ;\qquad {\bar d_{\,m}}\,\bar\phi\,=\,0\ ,
\ee
as well as  the ``inside-out" constraints

\be
\bar d_m^{}\,\bar d_n^{}\,\phi~=~\frac{1}{ 2}\,\epsilon_{mnpq}^{}\,d^p_{}\,d^q_{}\,\bar\phi\ ,
\ee
\be
 d^m_{}\, d^n_{}\,\bar\phi~=~\frac{1}{ 2}\,\epsilon^{mnpq}_{}\,\bar d_p^{}\,\bar d_q^{}\,\phi\ .
\ee
The Yang-Mills action is then simply

\be\int d^4x\int d^4\theta\,d^4 \bar \theta\,{\cal L}\ ,
\ee
where

\bea
{\cal L}&=&-\bar\phi\,\frac{\Box}{\partial^{+2}}\,\phi
~+\frac{4g}{3}\,f^{abc}_{}\,\Big(\frac{1}{\partial^+_{}}\,\bar\phi^a_{}\,\phi^b_{}\,\bar\partial\,\phi^c_{}+{\rm complex~conjugate}\Big)\cr
&&-g^2f^{abc}_{}\,f^{ade}_{}\Big(\,\frac{1}{\partial^+_{}}(\phi^b\,\partial^+\phi^c)\frac{1}{\partial^+_{}}\,(\bar \phi^d_{}\,\partial^+_{}\,\bar\phi^e)+\frac{1}{2}\,\phi^b_{}\bar\phi^c\,\phi^d_{}\,\bar\phi^e\Big)\ .
\eea
Grassmann integration  is normalized so that $\int d^4\theta\,\theta^1\theta^2\theta^3\theta^4=1$, and $f^{abc}$ are the structure functions of the  Lie algebra.

\subsection{SuperPoincar\'e Algebra}
 The generators of the Poincar\'e algebra are given by the four momenta

\be
p^-_{}~=~-i\frac{\partial\bar\partial}{\partial^+_{}}\ ,\qquad p^+_{}~=~-i\,\partial^+_{}\ ,\qquad p~=~-i\,\partial\ ,\qquad \bar p~=~-i\,\bar\partial\ ,
\ee
using light-cone coordinates and no interactions; it also contains the kinematical transverse space rotations

\be
j~=~x\,\bar\partial-\bar x\,\partial+ S^{12}_{}\ ,
\ee
where the little group helicity generator is 

\be
S^{12}_{}~=~ \,\frac{1}{ 2}\,(\,{\theta^p}\,{{\bar \partial}_p}\,-\,{{\bar \theta}_p}\,{\partial^p}\,)\,+\frac{i}{4\sqrt{2}\,\partial^+}\,(\,d^p\,\bar d_p-\bar d_p\,d^p\,)\ .
\ee
This form is slightly different from the original  expression of ref.  \cite{CHALMERS1} through the last term, which acts as a helicity counter.  It also ensures that the chirality constraints are preserved

\be
[\,j\,,\,d^m_{}\,]~=~[\,j\,,\,\bar d^{}_m\,]~=~0\ .
\ee
Under an infinitesimal $SO(2)$ transformation, this generator acts as a differential operator on the chiral superfield

\be
\delta\,\phi~=~i\,\omega\,j\,\phi\ ,\qquad \delta\,\bar \phi~=~-i\,\omega\,j\,\bar \phi\ .
\ee
The other kinematical generators are 

\be
j^+_{}~=~i\, x\,\partial^+_{}\ ,\qquad \bar j^+_{}~=~i\,\bar x\,\partial^+_{}\ .
\ee
The rest of the generators must be specified separately for chiral and antichiral fields. Acting on $\phi$, we have 

\be
 j^{+-}_{}~=~i\,x^-_{}\,\partial^+_{}-\frac{i}{2}\,(\,\theta^p_{}\bar\partial^{}_p+\bar\theta^{}_p\,\partial^p_{}\,)\ ,
\ee
chosen so as to  preserve the chiral combination
\be
[\,j^{+-}_{}\,,\,y^-_{}\,]~=~-i\,y^-_{}\ ,
\ee
and its commutators  with the chiral derivatives

\be
[\,j^{+-}_{}\,,\,d^m_{}\,]~=~\frac{i}{2}\,d^m_{}\ ,\qquad [\,j^{+-}_{}\,,\,\bar d_m^{}\,]~=~\frac{i}{2}\,\bar d^{}_m\ ,
\ee
 preserve chirality. Similarly the dynamical boosts are 

\bea
j^-_{}&=&i\,x\,\frac{\partial\bar\partial}{\partial^+_{}} ~-~i\,x^-_{}\,\partial~+~i\,\Big( \theta^p_{}\bar\partial^{}_p\,+\frac{i}{4\sqrt{2}\,\partial^+}\,(\,d^p\,\bar d_p-\bar d_p\,d^p\,)\Big)\frac{\partial}{\partial^+_{}}\,\ ,\cr 
\bar j^-_{}&=&i\,\bar x\,\frac{\partial\bar\partial}{\partial^+_{}}~ -~i\,x^-_{}\,\bar\partial~+~ i\,\Big(\bar\theta_p^{}\partial_{}^p+\frac{i}{4\sqrt{2}\,\partial^+}\,(\,d^p\,\bar d_p-\bar d_p\,d^p\,)\,\Big)\frac{\bar\partial}{\partial^+_{}}\,\ .
\eea
They do not commute with the chiral derivatives, 

\be
[\,j^{-}_{}\,,\,d^m_{}\,]~=~\frac{i}{2}\,d^m_{}\,\frac{\partial}{\partial^+_{}}\ ,\qquad [\,j^{-}_{}\,,\,\bar d_m^{}\,]~=~\frac{i}{2}\,\bar d_m^{}\,\frac{\partial}{\partial^+_{}}\ ,
\ee
but do not change  the chirality of the fields on which they act.  They   satisfy the Poincar\'e algebra, in particular

\be
[\,j_{}^-\,,\,\bar j^+_{}\,]~=~-i\,j^{+-}_{}-j\ ,\qquad [\,j^-_{}\,,\,j^{+-}_{}\,]~=~i\,j^{-}_{}\ .
\ee
On the light-cone,  supersymmetry breaks up into two types, kinematical and dynamical. The kinematical supersymmetries

\be
q^{\,m}_{\,+}=-{\partial^m}\,+\,\frac{i}{\sqrt 2}\,{\theta^m}\,{\partial^+}\ ;\qquad{{\bar q}_{\,+\,n}}=\;\;\;{{\bar \partial}_n}\,-\,\frac{i}{\sqrt 2}\,{{\bar \theta}_n}\,{\partial^+}\ ,
 \ee
satisfy

\be
\{\,q^{\,m}_{\,+}\,,\,{{\bar q}_{\,+\,n}}\,\}\,=\,i\,{\sqrt 2}\,{{\delta^m}_n}\,{\parp}\ .
\ee
and  anticommute with the chiral derivatives

\bea
\{\,q^{\,m}_{\,+}\,,\,{{\bar d}_n}\,\}\,=\,\{\,{d^m}\,,\,{{\bar q}_{\,+\,n}}\,\}\,=\,0\ .
\eea
The dynamical supersymmetries are obtained by boosting the kinematical ones

\be
{q}^m_{\,-}~\equiv~i\,[\,\bar j^-_{}\,,\,q^{\,m}_{\,+}\,]~=~\frac{\partial}{\partial^+_{}}\, q^{\,m}_{\,+}\ ,\qquad 
{\bar{q}}_{\,-\,m}^{}~\equiv~i\,[\, j^-_{}\,,\,\bar q_{\,+\,m}^{}\,]~=~\frac{\bar\partial}{\partial^+_{}}\, \bar q_{\,+\,m}^{}\ .
 \ee
They satisfy  the  free $N=4$ supersymmetry algebra

\be
\{\, {q}^m_{\,-}\,,\,{\bar{q}}_{\,-\,n}^{}\,\}~=~i\,\sqrt{2}\,\delta^{\,m}_{~~n}\,\frac{\partial\bar\partial}{\partial^+_{}}\ .
\ee
In the interacting theory, all dynamical generators will be altered by interactions, as discussed in ref. \cite{INTERACTION}. 
\section{Ten Dimensions}
The very compact formalism of the previous section was constructed for the $N=4$ theory in four dimensions. In this paper we  generalize this formalism to restore the theory in ten dimensions, without changing  the superfield, simply  by introducing  generalized derivative operators.  

First of all,  the  transverse light-cone  variables need to be generalized to eight. We stick to the previous notation, and introduce  the six extra coordinates   and their derivatives as antisymmetric bi-spinors

\be
{x^{m\,4}}\,=\,\frac{1}{\sqrt 2}\,(\,{x_{m\,+\,3}}\,+\,i\,{x_{m\,+\,6}}\,)\ ,\qquad 
{\partial^{m\,4}}\,=\,\frac{1}{\sqrt 2}\,(\,{\partial_{m\,+\,3}}\,+\,i\,{\partial_{m\,+\,6}}\,)\ ,
\ee
for $m\ne 4$, and their complex conjugates 

\be{{\bar x}_{pq}}\,=\,\frac{1}{2}\,{\epsilon_{pqmn}}\,{x^{mn}}\ ;\qquad{{\bar \partial}_{pq}}\,=\,\frac{1}{2}\,{\epsilon_{pqmn}}\,{\partial^{mn}}\ .
\ee
Their derivatives satisfy

\be
{{\bar \partial}_{mn}}\,{x^{pq}}\,~=~\,(\,{{\delta_m}^p}\,{{\delta_n}^q}\,-\,{{\delta_m}^q}\,{{\delta_n}^p}\,) \ ;\qquad 
{\partial^{mn}}\,{{\bar x}_{pq}}~=~(\,{{\delta^m}_p}\,{{\delta^n}_q}\,-\,{{\delta^m}_q}\,{{\delta^n}_p}\,) \ ,
\ee
and

\be
{{\partial}^{mn}}\,{x^{pq}}~=~\frac{1}{2}\,{\epsilon^{pqrs}}\,{{\partial}^{mn}}\,{{\bar x}_{rs}}~=~{\epsilon^{mnpq}}\ .
\ee
There are no modifications to be made to the chiral superfield, except for the  dependence  on the extra coordinates

\be
A(y)~=~A(x,\bar x,x^{mn}_{},\bar x_{mn}^{},y^-_{})\ ,~~etc... ~\ .
\ee 
These extra variables  will be acted on by new operators that generate the higher-dimensional symmetries.

\subsection{The SuperPoincar\'e Algebra in $10$ Dimensions}
The SuperPoincar\'e algebra needs to be generalized from the  form in ref. \cite{CHALMERS1}.  One starts with the construction of the $SO(8)$ little group 
using   the decomposition $SO(8)\supset SO(2)\times SO(6)$. The $SO(2)$ generator is the same; the $SO(6)\sim SU(4)$ generators are given by  

\bea
{J^m}_n\,&=&\,\frac{1}{ 2}\,(\,{x^{mp}}\,{{\bar \partial}_{pn}}\,-\,{{\bar x}_{pn}}\,{\partial^{mp}}\,)\,-\,{\theta^m}\,{{\bar \partial}_n}\,+\,{{\bar \theta}_n}\,{\partial^m}\,+\,\frac{1}{4}\,(\,{\theta^p}\,{{\bar \partial}_p}\,-\,{{\bar \theta}_p}\,{\partial^p}\,)\,{{\delta^m}_n}\cr
& &+ \frac{i}{2\sqrt{2}\,\partial^+}\,(\,d^m\,\bar d_n-\bar d_n\,d^m\,)+\frac{i}{8\sqrt{2}\,\partial^+}\,(\,d^p\,\bar d_p-\bar d_p\,d^p\,)\,{\delta^m}_{~n}\ .
\eea
The extra terms with the $d$ and $\bar d$ operators are not necessary for closure of the algebra. However they insure that the generators commute with the chiral derivatives. They satisfy the commutation relations

\bea
\Big[\,J\,,\,{J^m}_n\,\Big]~=~0\ ,\qquad 
\Big[\,{J^m}_n\,,\,{J^p}_q\,\Big]~=~{\delta^m}_q\,{J^p}_n-{\delta^p}_n\,{J^m}_q\ .
\eea
The remaining $SO(8)$ generators lie in the  coset  $SO(8)/(SO(2)\times SO(6))$ 

\bea
J^{pq}&=&x\,\partial^{pq}-x^{pq}\,\partial+\frac{i}{\sqrt 2}\,{\parp}\,{\theta^p}\,{\theta^q}-i\,{\sqrt 2}\,\frac{1}{\parp}\,{\partial^p}\,{\partial^q} +\frac{i}{\sqrt{2}\,\partial^+}\,d^p\,d^q\ ,\cr\cr
{{\bar J}_{mn}}&=&{\bar x}\,{{\bar \partial}_{mn}}-{{\bar x}_{mn}}\,{\bar \partial}+\frac{i}{{\sqrt 2}}\,{\parp}\,{{\bar \theta}_m}\,{{\bar \theta}_n}-i\,{\sqrt 2}\,\frac{1}{\parp}\,{{\bar \partial}_m}\,{{\bar \partial}_n}+\frac{i}{\sqrt{2}\,\partial^+}\,\bar d_m\,\bar d_n\ .
\eea
All   $SO(8)$ transformations are specially constructed so as not to mix chiral and antichiral superfields, 

\be
\,[\,{J^{mn}}\,,\,{{\bar d}_p}\,]\,~=~0\ ;\qquad \,[\,{{\bar J}_{mn}}\,,\,{d^p}\,]\,~=~0\ ,
\ee
and satisfy the $SO(8)$ commutation relations 

\beas
\Big[\,J\,,\,J^{mn}_{}\,\Big]&=&J^{mn}_{}\ ,\qquad \Big[\,J\,,\,\bar J_{mn}^{}\,\Big]~=~-\bar J_{mn}^{}\ ,\cr 
&&\cr
\Big[\,{J^m}_n\,,\,J^{pq}_{}\,\Big]&=&{\delta^q}_n\,J^{mp}_{}-\,{\delta^p}_n\,J^{mq}_{}\ ,\qquad
\Big[\,{J^m}_n\,,\,\bar J^{}_{pq}\,\Big]~=~{\delta^m}_q\,\bar J_{np}^{}\,-\,{\delta^m}_p\,\bar J_{nq}^{}\ ,\cr
&&\cr
{\Big[}\,{J^{mn}}\,,\,{{\bar J}_{pq}}\,{\Big ]}&=&{{\delta^m}_p}{{J^n}_q}\,+\,\,{{\delta^n}_q}{{J^m}_p}\,-\,{{\delta^n}_p}{{J^m}_q}\,-\,\,{{\delta^m}_q}{{J^n}_p}\,-\,(\,{{\delta^m}_p}\,{{\delta^n}_q}\,-\,{{\delta^n}_p}\,{{\delta^m}_q}\,)\,J\ .
\eeas
Rotations between the $1$ or $2$ and $4$ through $9$ directions induce on the chiral fields the changes

\be
\delta\,\phi~=~\Big(\,\frac{1}{2}\,\omega^{}_{mn}\,J^{mn}_{}+ \frac{1}{2}\,\bar\omega_{}^{mn}\,\bar J_{mn}^{}\,\Big)\,\phi\ ,
\ee
where complex conjugation is like duality

\be
{\bar \omega}_{pq}\,=\,\frac{1}{2}\,\epsilon^{}_{mnpq}\,\omega_{}^{mn}\ .
\ee
For example, a  rotation in the $1-4$ plane through an angle $\theta$ corresponds to taking $\theta=\omega_{14}=\omega_{23}$ ($ =\omega^{23}=\omega^{14}$ by reality), all other components being zero. Finally, we verify that the kinematical supersymmetries are duly rotated by these generators

\be
\,[\,{J^{mn}}\,,\,\bar q_{+\,p}^{}\,]~=~{{\delta^n}_p}\,{q^m_+}\,-\,{\delta^m}_p\,q^n_+ \ ;\qquad 
[\,{\bar J}_{mn}\,,\,{q^p_+}\,]~=~{{\delta_n}^p}\,{\bar q}_{+\,m}^{}\,-\,{\delta_m}^p\,{\bar q}_{+\,n}^{}\ .
\ee
We  now use the $SO(8)$ generators to construct the SuperPoincar\'e generators

\bea
J^+_{}&=&i\,x\,\partial^+_{}\ ;\qquad \bar J^{+}_{}~=~i\,\bar x\,\partial^+_{} \cr &&\cr
J^{+\,mn}_{}&=&i\,x^{mn}_{}\,\partial^+_{}\ ; \qquad 
\bar J^+_{~~mn}~=~i\,\bar x^{}_{mn}\,\partial^+_{}\ . 
\eea
The dynamical boosts are now 

\bea
J^-_{}&=&i\,x\,\frac{\partial\bar\partial\,+\,{\frac {1}{4}}\,{{\bar \partial}_{pq}}\,{\partial^{pq}}}{\partial^+_{}} ~-~i\,x^-_{}\,\partial+i\,{\frac { \partial}{\parp}}\,\Big\{\,{ \theta}^m\,{\bar\partial_m}~+~\frac{i}{4\sqrt{2}\,\parp}\,(d^p\,\bar d_p-\bar d_p\,d^p)\,\Big\}-\cr
&&~~~~~~~~~~~~~-~  {\frac {1}{4}}\,{\frac {\bar\partial_{pq}}{\parp}}\,{\biggl \{}\,\frac{\parp}{{\sqrt 2}}\,{{ \theta}^p}\,{{ \theta}^q}\,-\,\,\frac{\sqrt 2}{\parp}\,{{\partial}^p}\,{{ \partial}^q}+\frac{1}{\sqrt{2}\parp}d^p\,d^q\,\,{\biggr \}}\ ,
\eea
and its conjugate

\bea
\bar J^-_{}&=& i\,\bar x\,\frac{\partial\bar\partial\,+\,{\frac {1}{4}}\,{{\bar \partial}_{pq}}\,{\partial^{pq}}}{\partial^+_{}}-~i\,x^-_{}\,\bar\partial~+~i\,{\frac {\bar \partial}{\parp}}\,\Big\{\,{\bar \theta}_m\,{\partial^m}~+~\frac{i}{4\sqrt{2}\,\parp}\,(d^p\,\bar d_p-\bar d_p\,d^p)\,\Big\}-~\cr
 &&~~~~~~~~~~~~~-~ {\frac {1}{4}}\,{\frac {\partial^{pq}}{\parp}}\,{\biggl \{}\,\frac{\parp}{{\sqrt 2}}\,{{\bar \theta}_p}\,{{\bar \theta}_q}\,-\,\,\frac{\sqrt 2}{\parp}\,{{\bar \partial}_p}\,{{\bar \partial}_q}+\frac{1}{\sqrt{2}\parp}\bar d_p\,\bar d_q\,{\biggr \}}\ .
\eea
The others are obtained by using the $SO(8)/(SO(2)\times SO(6))$  rotations

\be
J^{-\,mn}_{}~=~[\,J^-_{}\,,\,J^{mn}_{}\,]\ ;\qquad 
\bar J^-_{~~mn}~=~[\,\bar J^-_{}\,,\,\bar J^{}_{mn}\,]\ .
\ee
We do not show their explicit forms as they are too cumbersome. 
The four supersymmetries in four dimensions turn into one supersymmetry in ten dimensions. In our notation, the kinematical supersymmetries  $q^n_+$ and $\bar q^{}_{+n}$, are assembled into one $SO(8)$ spinor. The dynamical supersymmetries are obtained by boosting 

\be
i\,[\,{\bar J}^-\,,\,q_+^m\,]~\equiv~{\cal Q}^m_{}\ ,\qquad 
i\,[\,J^-\,,\,\bar q_{+\,m}\,]~\equiv~\nbar{\cal Q}_m^{}\ ,
\ee
where 

\bea
{\cal Q}^m_{}&=&{\frac {\bar \partial}{\parp}}\,{{q_+}^m}\,+\,{\frac {1}{2}}\,{\frac {\partial^{mn}}{\parp}}\,{{\bar q}_{+\,n}}\ ,\cr
&&\cr
\nbar{\cal Q}_m^{}&=&\frac {\partial}{\parp}}\,{{\nbar q}_{+\,m}}\,+\,{\frac{1}{2}}\,{\frac {{\nbar \partial}_{mn}}{\parp}\, q^{~n}_{\,+}\ .
\eea
They satisfy the supersymmmetry algebra

\be
\{\,{\cal Q}^{\,m}_{}\,,\,\nbar {\cal Q}^{}_{\,n}\,\}~=~i\,\sqrt{2}\,\delta^m_{~n}\,\frac{1}{\parp}
\,\Big(\partial\,{\nbar \partial}\,+\,\frac{1}{4}\,{\nbar \partial}_{pq}\,\partial^{pq}\,\Big)\ ,
\ee
and can be  obtained from one another by $SO(8)$ rotations, as 
\be
\frac{1}{2}\,{\epsilon_{pqmn}}\;[\,J^{pq}\,,\,{\cal Q}^{\,m}\,]\;=\;~4\,{\nbar {\cal Q}}_{\,n}\ ,
\ee
while 
\be
[\,\bar J_{pq}\,,\,{\cal Q}^{\,m}\,]~=~0\ .
\ee

Note also that 

\be
\{\,{\cal Q}^{\,m}_{}\,,\,q_+^n\,\}~=~\frac{i}{\sqrt 2}\, \partial^{mn}\ .
\ee
We thank the anonymous referee for suggesting that we look at central charges in this framework. 
It provides a simple method to introduce the central charges germane to the four-dimensional theory: 
the six-dimensional derivatives $\partial^{mn}$, are simply replaced by 
c-numbers $Z^{mn}$, thus yielding the massive supersymmetry algebra in 
four dimensions with six central charges,
\beas
{Z^{mn}}\,=\,{\frac {1}{2}}\,{\epsilon^{mnpq}}\,{{\nbar Z}_{pq}}\ .
\eeas
However, relaxing this duality condition, gives us all $12$ central charges of the $N=4$ Yang-Mills Theory, thus indicating that both varieties of central charges have a common origin in the light-cone gauge formulation.

Starting from these results for the $N=1$ SuperPoincar\'e generators for the free theory, we proceed to  build 
the interacting theory in ten dimensions. 

%%%%%%%%%%%%%%%%%%%%%%%%%%%%%%%%%%%%%%%%%%%%%%%%%%%%%%%%%%%%%%%%%%%%%%%%%%%%

\subsection{The Generalized Derivatives}
The cubic interaction in the  $N=4$ Lagrangian contains explicitly the  derivative operators $\partial$ and $\bar\partial$. To achieve covariance in ten dimensions, these must be generalized. We propose the following operator 
\be
{\nbar \nabla}~\equiv~{\bar \partial}\,+\,\frac{i\,\alpha}{4\,\sqrt 2\,\partial^+}\,{{\bar d}_p}\,{{\bar d}_q}\,\partial^{pq} \ ,
\ee
which naturally incorporates the rest of the derivatives $\partial^{pq}$, with   $\alpha$ as an arbitrary parameter. After some algebra, we find that $\nbar\nabla$ is covariant under $SO(8)$ transformations. We define its rotated partner as 

\be
{\nabla_{}^{mn}}~\equiv~{\Big[}\,{\nbar \nabla}\,,\,{J^{mn}}\,{\Big]}\ , 
\ee
where

\be
{\nabla_{}^{mn}}~=~{\partial^{mn}_{}}\, -\,\frac{i\,\alpha}{4\,\sqrt 2\,\parp}\,{{\bar d}^{}_r}\,{{\bar d}^{}_s}\,{\epsilon_{}^{mnrs}}\,\partial\ .
\ee
If we apply to it the inverse transformation, it goes back to the original form 

\be
\Big [\,{\nbar J}_{pq}\,,\,\nabla^{mn}\,\Big] \,~=~\,(\,{\delta_p}^m\,{\delta_q}^n\,-{\delta_q}^m\,{\delta_p}^n\,)\,\nbar \nabla\ ,
\ee
and these operators transform under   $SO(8)/(SO(2)\times SO(6))$, and $SO(2)\times SO(6)$  as  the components of an $8$-vector.  

We introduce the conjugate operator $\nabla$ by requiring that 

\be
\nabla\,\bar\phi~\equiv~ \nbar{(\nbar\nabla\,\phi)}\ ,
\ee
with  

\be
{ \nabla}~\equiv~{\partial}\,+\,\frac{i\,\alpha}{4\,\sqrt 2\,\partial^+}\,{{ d}^p}\,{{ d}^q}\,\bar\partial_{pq} \ .
\ee
Define 

\be
{\nbar \nabla}_{\;mn}~{\equiv}~{\Big[}\,{{\nabla}}\,,\,{{\bar J}_{mn}}\,{\Big]}\ ,
\ee
which is given by 
\be
{\nbar \nabla^{}_{mn}}~=~{\bar\partial_{mn}^{}}\, -\,\frac{i\,\alpha}{4\,\sqrt 2\,\parp}\,{{ d}^r_{}}\,{{ d}_{}^s}\,{\epsilon_{mnrs}^{}}\,\bar \partial\ .
\ee
We then  verify that 

\be
\Big [\,{ J}^{mn}\,,\,\nbar\nabla_{pq}\,\Big] \,~=~\,(\,{\delta_p}^m\,{\delta_q}^n\,-{\delta_q}^m\,{\delta_p}^n\,)\, \nabla\ .  
\ee
The value of the  parameter $\alpha$ is fixed by the invariance of the cubic interaction. 

\subsection{Invariance of the Action}
The kinetic term is trivially made $SO(8)$-invariant by including the six extra transverse derivatives in the d'Alembertian. The quartic interactions are  obviously invariant since they  do not contain any transverse derivative operators. Hence we need only consider the cubic vertex.

The point of our paper is that to achieve covariance in ten dimensions, it suffices to replace the transverse $\partial$ and $\bar\partial$ by $\nabla$ and $\nbar\nabla$, respectively. We propose the  new cubic interaction term  

\be
\frac{4g}{3}\,f^{abc}_{}\int d^{10}x\,\int\, d^4\,\theta\,d^4\,\bar\theta\,\Big(\frac{1}{\partial^+_{}}\,\bar\phi^a_{}\,\phi^b_{}\,\nbar\nabla\,\phi^c_{}+{\rm complex~conjugate}\Big)
\ee
Since it is obviously invariant under $SO(6)\times SO(2)$, we need only consider the coset variations. On the chiral superfield, using the chiral constraints,

\be
{\delta_{J}}\,{\phi}~\equiv~{{\nbar \omega}_{mn}}\,J^{mn}\,{\phi}\,=\,i\,{\sqrt 2}\;\;{{\nbar \omega}_{mn}}\,{\parp}\,{\theta^m}\,{\theta^n}\,{\phi}
\ee

\be
{\delta_{\bar J}}\,{\phi}~\equiv~{{\omega}^{pq}}\,{{\nbar J}_{pq}}\,{\phi}\,=\,{{\omega}^{pq}}\,{\biggl \{}\,\frac{i}{{\sqrt 2}}\,{\parp}\,{{\bar \theta}_p}\,{{\bar \theta}_q}\,-\,i\,{\sqrt 2}\,\frac{1}{\parp}\,{{\bar \partial}_p}\,{{\bar \partial}_q}\,+\,{\frac{i}{\sqrt 2}}\,{1\over \parp}\,{{\nbar d}_p}\,{{\nbar d}_q}\,{\biggr \}}\,{\phi}\ .
\ee
We list the conjugate relations for completeness

\be
{\delta_{\bar J}}\,{\nbar \phi}~=~{{\omega}^{pq}}\,{{\nbar J}_{pq}}\,{\nbar\phi} ~=~i\,{\sqrt 2}\;\;{{\omega}^{pq}}\,{\parp}\,{{\bar \theta}_p}\,{{\bar \theta}_q}\,{\nbar \phi}\ ,
\ee

\be
{\delta_{J}}\,{\nbar \phi}\;=\;{{\nbar \omega}_{mn}}\,J^{mn}\,{\nbar \phi}\,=\,{{\nbar \omega}_{mn}}\,{\biggl \{}\,\frac{i}{\sqrt 2}\,{\parp}\,{\theta^m}\,{\theta^n}\,-\,i\,{\sqrt 2}\,\frac{1}{\parp}\,{\partial^m}\,{\partial^n}\,+\,{\frac{i}{\sqrt 2}}\,{1\over \parp}\,{d^m}\,{d^n}\,{\biggr \}}\,{\nbar \phi} \ .
\ee
The variations of the generalized derivative and its conjugate are given by 

\be
{\delta_{J}}\,{\nbar \nabla}\,=\,{{\nbar \omega}_{mn}}\,[\,J^{mn}\,,\,{\nbar \nabla}\,]\,=\,-\,{{\nbar \omega}_{mn}}\,{\nabla^{mn}} \ ,\\\\
\ee
\be
{\delta_{\bar J}}\,{\nbar \nabla}\,=\,{\omega^{pq}}\,[\,{{\nbar J}_{pq}}\,,\,{\nbar \nabla}\,]\,=\,{\frac {i\,{\alpha}}{2\,{\sqrt 2}}}\,{\omega^{pq}}\;{{\bar d}_p}{{\bar d}_q}\,{\frac {\nbar \partial}{\parp}}\ .
\ee
Invariance under $SO(8)$ is checked by doing  a $\delta_J$ variation  on the  cubic vertex, {\it including} its complex conjugate.  In terms of 

\be
{{\cal V}}\;+\;{\nbar {\cal V}}~\equiv~{{\mathit {f}}_{abc}}\,{\int}\;\Big({1\over \parp}\;{{\nbar \phi}^a}\;\;{\phi^b}\;\;{\nbar \nabla}\;{\phi^c}~+~
{1\over \parp}\;{\phi^a}\;\;{{\nbar \phi}^b}\;\;{\nabla}\;{{\nbar \phi}^c}\Big)\ ,
\ee
explicit calculations yield

\bea
{\delta_{J}}\,{\cal V}~=~ {{\mathit {f}}_{abc}}\,\;{{\nbar \omega}_{mn}}\,{\int}\;\Big(\,(\;{\frac {\alpha}{2}}\,-\,1\,)\,{\frac{1}{\parp}}\,{{\nbar \phi}^a}\;{\phi^b}\;{\partial^{mn}}\,{\phi^c}\;+\;{\frac {i\,\alpha}{2\,{\sqrt 2}}}\,{\frac{1}{\parp}}\,{{\nbar \phi}^a}\;{\phi^b}\;{d^m}\,{d^n}\,{\frac {\partial}{\parp}}\,{{\nbar \phi}^c}\,\Big)\ ,
\eea
and

\bea
{\delta_{J}}\,{\nbar {\cal V}}&=&\nbar{\delta_{\bar J}\,{\cal V}}\nonumber\\
&=&{{\mathit {f}}_{abc}}\,\;{{\nbar \omega}_{mn}}\,{\int}\Big(-{\frac {\alpha}{2}}\,{{\phi}^a}\;\frac{1}{\parp}\,{{\nbar \phi}^b}\;{\partial^{mn}}\,{\phi^c}\;+\;{\frac {i}{\sqrt 2}}\,{\phi^a}\;{\frac{1}{\parp}}\,{{\nbar \phi}^b}\;{d^m}\,{d^n}\,{\frac {\partial}{\parp}}\,{{\nbar \phi}^c} \nonumber \\
& &~~~~~~~~~~~~~~~~~~~~~~~~~~+\;{\frac {i\,{\alpha}}{2\,{\sqrt 2}}}\,\,{\frac{1}{\parp}}\,{\phi^a}\;{{\nbar \phi}^b}\;{d^m}\,{d^n}\,{\frac {\partial}{\parp}}\,{{\nbar \phi}^c}\,\Big)
\eea
Adding the two yields the final result

\bea
{\delta_{J}}\,\big({\cal V}\;+\;{\nbar{\cal V}}\big)&=&(\,{\alpha}\,-\,1\,)\,{{\mathit {f}}_{abc}}\,\;{{\nbar \omega}_{mn}}\;\times \cr&&\cr
& &\times{\int}\;{\biggl (}{\frac{1}{\parp}}\,{{\nbar \phi}^a}\;{\phi^b}\;{\partial^{mn}}\,{\phi^c}\;+\;{\frac {i}{\sqrt{2}\,\parp}}\,{{\nbar \phi}^a}\;{\phi^b}\;{d^m}\,{d^n}\,{\frac {\partial}{\parp}}\,{{\nbar \phi}^c}{\biggr )}\ .
\eea
The cubic vertex is $SO(8)$ invariant if $\alpha=1$, and the generalized derivative is totally determined

\be
{\nbar \nabla}~=~{\bar \partial}\,+\,\frac{i}{4\,\sqrt 2\,\partial^+}\,{{\bar d}_p}\,{{\bar d}_q}\,\partial^{pq} \ .
\ee
To be sure, we have checked invariance by performing the Grassmann integrations, and looking at the components.  

To obtain this result, we have used  the antisymmetry of the structure functions, the chiral constraints,  the ``inside-out"  constraints, and performed 
integrations by parts on the coordinates and  Grassmann variables.  In particular, using  the relation between the chiral field and its conjugate,  implied by the ``inside out" constraints

\be
{\nbar \phi}~=~\frac{1}{2\cdot 4!}\,\epsilon_{}^{pqmn}\,{{\bar d}_p}\,{{\bar d}_q}\,{{\bar d}_m}\,{{\bar d}_n}\,\frac{1}{{\parp}^2}\,{\phi}\ ,
\ee
we deduce two magical identities  

\be
{{\mathit {f}}_{abc}}\,{\int}\;{\frac {1}{{\parp}^2}}\,{{\nbar \phi}^a}\;{\phi^b}\;{\bar \partial}\,{\phi^c}~=~0\ ,
\ee
\be
{{\mathit {f}}_{abc}}\,{\int}\;{\frac {1}{\parp}}\,{{\nbar \phi}^a}\;{\frac {1}{\parp}}\,{\phi^b}\;{d^m}\,{d^n}\,{\partial}\,{{\nbar \phi}^c}~=~0\ .
\ee
In this light-cone form, the Lorentz invariance in ten dimensions is automatic once the little group invariance has been established. We have therefore shown  ten-dimensional invariance, since the quartic term does not need to be changed.

\section{Conclusions}
We have explicitly constructed the light-cone action of ten-dimensional SuperYang-Mills, starting from its four-dimensional realization. Crucial to the construction is the generalized derivative whose components form an $SO(8)$ vector. Although its geometrical meaning is not entirely clear to us, it suggests that this action has further hidden symmetries, 
an issue  we hope to come back to in the  future. 

As we stated in the introduction, this oxidation of the light-cone formulation of the $N=4$ theory should naturally lead us to a similar phenomenon for $N=8$ supergravity in four dimensions. In that case, we have to add in the seven missing coordinates and the new derivatives should look very much like the one in (63), with $SO(7)$ substituted for $SO(6)$.  In a future publication\cite{ABR}, we will show how it leads along similar lines to $N=1$ Supergravity in eleven dimensions.

\end{document}